\documentclass{SCGE}
\usepackage[colorlinks,linkcolor=blue,citecolor=blue,urlcolor=blue]{hyperref}

\usepackage[latin1]{inputenc}
\usepackage{rotating}
\usepackage{footnote}
\newcolumntype{C}[1]{>{\centering\let\newline\\\arraybackslash\hspace{0pt}}m{#1}}
\usepackage[math]{blindtext}
\usepackage{times}
\usepackage{newtxtext,newtxmath}
\usepackage{graphicx}	
\usepackage{amsmath}	
\usepackage{amssymb}	
\usepackage{xspace}
\usepackage{bigints}
\usepackage{siunitx}
\usepackage{amsthm}
\usepackage{mathtools}
\usepackage[normalem]{ulem}
\usepackage{float}
\usepackage{color}
\usepackage[sort&compress,numbers]{natbib}

\usepackage{CJK}

\newcommand{\gray}{$\gamma$-ray~}
\newcommand{\grays}{$\gamma$-rays~}

\let\citedash\relax
\makeatletter \providecommand{\citedash}{\hbox{-}\penalty\@m}
\makeatother
\begin{document}

\begin{picture}(0,0){\rm
\put(0,-20){\makebox[160truemm][l]{\bf {\sanhao\raisebox{2pt}{.}}
Article  {\sanhao\raisebox{1.5pt}{.}}}}}
\put(0,-34){\jiuwuhao {\textcolor[rgb]{0.5,0.5,0.5}{\sf 
}}}
\end{picture}

\def\bm{\boldsymbol}

\def\dl{\displaystyle}
\def\du{\end{document}}
\def\d{{\rm d}}
\def\e{{\rm e}}
\def\i{{\rm i}}

\Year{2021} %
\Month{XXXX} %
\Vol{xx} %
\No{x} %
\BeginPage{1} %
\AuthorMark{{\rm R Z. YANG,  B. LIU} }  
\DOI{} 
\ArtNo{000000}

\title[radial profile of \gray sources]{On the surface brightness radial profile of the extended \gray sources }

\author[1,2,3]{Rui-zhi Yang}{}%
\author[1,2,3,4]{Bing Liu*}{}
\footnote{*Corresponding author: lbing@ustc.edu.cn}
\address[{\rm1}]{Department of Astronomy, School of Physical Sciences, University of Science and Technology of China, Hefei, Anhui 230026, China;}
\address[{\rm2}]{CAS Key Labrotory for Research in Galaxies and Cosmology, University of Science and Technology of China, Hefei, Anhui 230026, China;}
\address[{\rm3}]{School of Astronomy and Space Science, University of Science and Technology of China, Hefei, Anhui 230026, China;}
\address[{\rm4}]{Key Laboratory of Modern Astronomy and Astrophysics (Nanjing University), Ministry of Education, Nanjing 210093, China}

\maketitle \vspace{-3.5mm}{\footnotesize\begin{center} Received Month date, Year; accepted Month date, Year
\end{center}}\vspace*{-5mm}

\begin{center}
\rule{16.5cm}{0.4pt}
\parbox{16.5cm}
{\begin{abstract}
The morphology of the extended \gray source is governed by the propagation process of parent relativistic particles. In this paper, we investigate the surface brightness radial profile of extended \gray sources illuminated by  cosmic ray protons and electrons, considering the radiation mechanisms,  projection effects, and the response  of instruments. We found that the parent particle species and the propagation process can cause considerable differences in the observed radial profiles. Thus, the surface brightness profile can be used as a unique tool to identify the radiation mechanism and the propagation process of the parent particles.  In addition, we also discuss the possible implications regarding the latest discoveries from very/ultra-high energy \gray instruments like LHAASO and HAWC.

\end{abstract}}
\end{center}\vspace*{-0.6cm}

\begin{center}
\parbox{16.5cm}
{\bf\jiuhao Cosmic rays, $\gamma$-ray sources, diffuse
emission}
\end{center}

\begin{center}
{}
\end{center}

\textwidth=178truemm \textheight=236truemm

\wuhao\vspace*{1.5mm}

\renewcommand{\baselinestretch}{1.08} \baselineskip 12.2pt\parindent=10.8pt


\section{Introduction}
\label{sec1}
Thanks to the advances in the observational instruments,  observational \gray astronomy has achieved  unprecedented accuracy and sensitivities. Hundreds of extended sources have been discovered in the GeV to multi-TeV energy range \cite{4fgl,hgps,3hwc}. 
Their surface brightness radial profiles 
have been measured to study the radiation mechanism and the acceleration and propagation of relativistic particles. Recently revealed $1/r$ profiles in the vicinity of the Galactic Center (GC) and several young massive clusters (YMCs) indicate a continuous injection of cosmic ray (CR) protons from the central source to the surrounding CR "cocoons" \cite{hess2016pev,aharonian19}. Moreover, HAWC has detected halo structures near old pulsars, called "TeV halos", and the diffusion coefficient of the relativistic electrons therein derived from the brightness profiles is orders of magnitude less than the average value in the Galactic plane \cite{hawc_geminga}.
Both CR cocoons and TeV halos are identified through the radial profiles of the parent CRs, which are derived from the \gray profiles.  Thus, it is very interesting and important to investigate  the 
radial distribution of the extended \gray sources in detail, especially for the next generation \gray instruments such as LHAASO \cite{lhaaso2019} and CTA \cite{Actis2011},  since they can play a crucial role in identifying the particle accelerator and pinning down the propagation mechanisms.

Spatial distributions of \gray emissions are determined by  the distribution of their parent CRs (both electrons and protons) and the distribution of target fields, such as gas and interstellar radiation fields (ISRF). The gas distributions can be derived from molecular and atomic line observations, such as CO \cite{Dame2001co} and HI \cite{lab_survey}, and the infrared observations of  dust emissions \cite{planck_darkgas}.  Meanwhile, the ISRF can be derived by solving the radiation transfer process in our Galaxy \cite{porter19,Popescu2017radif}, and by star counting in dense regions \citep[see, e.g., ][]{fermi_cygnus}. Therefore, in principle, the \gray spatial distributions can be used to determine the distributions of the parent CRs. 
And from a theoretical point of view, the distribution of parent CRs mainly depends on the propagation process, including diffusion, advection, energy loss, etc, which have been studied extensively \citep[see, e.g.,][]{Atoyan1995,Atoyan2000,Strong2007review}. However, it should be mentioned that these distributions are usually calculated in a 3-dimensional (3D) configuration space, while the observed brightness is projected into a 2-dimensional (2D) surface. Thus, a careful treatment of such a projection process is also needed. In this paper, we assumed a spherical symmetry of the 3D distributions in our calculation. Such an assumption is valid in large scales for CR propagation, ISRF and gas distributions, such as in the case of Geminga \citep{hawc_geminga} and  GC \citep{hess2016pev}. However, at a smaller scale, both the gas distribution and CR propagation \citep{liu19ani} can be anisotropic, and such effects may need further investigations.  Furthermore, the projected profile should also be convolved with the instrument response (mainly point spread function (PSF) for spatial analysis) to compare with the observables. 
In this paper, we considered all these issues and calculated the  \gray profiles in different astrophysical scenarios and investigated how to distinguish them with the current and forthcoming instruments. 

This paper is organized as follows. In Section~\ref{sec2}, we derive the CR distribution near accelerators in various astrophysical cases. In Section~\ref{sec3}, we calculate the projection effects by adopting the Abel transformation.  We investigate the effects of instruments and  discuss a special case in which ballistic propagation is considered  in Section~\ref{sec4} and~\ref{sec5}, respectively. In Section~\ref{sec6}, the possible link between the extended sources and diffuse emissions is presented. Finally, we discuss the possible implications of our studies in Section~\ref{sec7}.

\section{CR distributions near accelerators}
\label{sec2}
The injection and propagation processes of CRs from their acceleration sites determine their  spatial distributions in the interstellar medium. Such processes have been intensively investigated in previous studies \citep[e.g., ][]{Atoyan1995,Aharonian1996AA}. Under the assumption of spherical symmetry, the diffusion equation with energy loss can be expressed as

\begin{equation}
\label{eq_dif}
    \frac{\partial F}{\partial t}=\frac{D}{R^2}\frac{\partial}{\partial R}R^2\frac{\partial F}{\partial t}+\frac{\partial}{\partial E}(PF)+Q,
\end{equation}

where F is the energy distribution of CR protons/electrons,  $Q$ represents the CR injection rate, $P$ is the energy-loss rate, $R$ represents the radial distance to the source, and $D$ is the diffusion coefficient of CRs. 
For the burst-like injection, the CR distribution at a given time $t$ after the injection can be expressed as the Green function of Eq.\ref{eq_dif},
\begin{equation}
\label{eq_green}
F_{\rm}(R,E,t)=\frac{Q(E_t)P(E_t)}{\pi^{3/2}P(E)R_{\rm diff}^3}  {\rm exp}\left(-\frac{R^2}{R_{\rm diff}^2}\right).
\end{equation}
Here $E_t$ is the initial energy of the particles that cool to energy $E$ at time $t$,  $R_{\rm diff}=2\sqrt{\Delta u}$ and $\Delta u(E,x) = \int^{x}_{E} \frac{D(E')}{P(E')}dE'$. The distribution is nearly uniform for $R<R_{\rm diff}$ and declines sharply for $R>R_{\rm diff}$. Therefore, in this case, the \gray emission near the source should have a good spatial correlation with the target fields such as gas and ISRFs. 

\begin{figure}[H]
  \centering
  \includegraphics[scale=0.8]{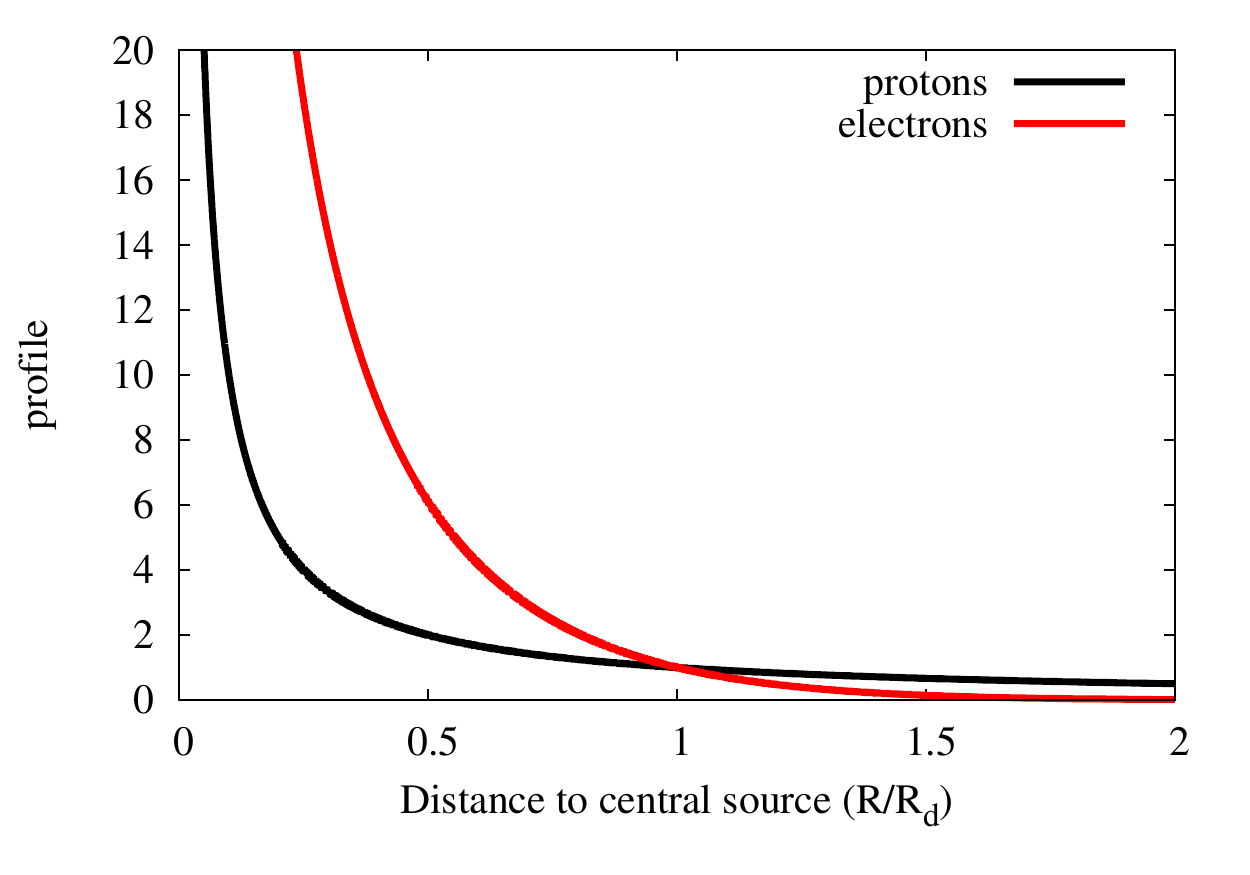}
  \caption{Distribution of CR protons and electrons near the acceleration site. }
   \label{fig:pro_3d}
\end{figure}

For continuous injections, the CR distribution is derived by integrating Eq.\ref{eq_green} over the injection history. For a constant injection spectrum $Q(E)$  during the time $t$, the integration  can be expressed as 
\begin{equation}
\label{eq_con}
F_{\rm con}(R,E,t) = \frac{Q(E){\rm erfc}(R/R_{\rm d})}{4\pi R D(E)},
\end{equation}
where  $R_{\rm d}$ is the diffusion length and can be approximated with $2 \sqrt{D t_{\rm \gamma}}$, where $t_{\rm \gamma}$ is the minimum of the total injection time $t$ and the cooling time $t_{\rm  c}$. For protons with kinematic energy above 1 GeV, the cooling effect can be neglected and $R_{\rm d}$ is merely determined by the injection time $t$. For regions near the source, $\sqrt{D t}$ is much larger than the region of interests in our study, thus t can be treated as infinity. In this case, the distribution scales as   $F_{\rm con}(R) \sim  \frac{1}{R D}$. 
However,  for CR electrons and protons below 1 GeV, the cooling effect must be considered in the propagation process. Thus, the distribution reads $F_{\rm con}(R) \sim \frac{{\rm erfc}(R/R_{\rm d})}{R D}$, where $R_{\rm d} \sim 2 \sqrt{D t_{\rm  c}}$. 
In this paper, we focus on the \gray emissions without considering protons below 1 GeV. We plotted the profile of these electrons and protons above 1 GeV in Fig.\ref{fig:pro_3d}, and we found that the distribution of electrons is steeper than that of protons.

\section{projection on a 2D surface}
\label{sec3}
Based on the CR distributions calculated in Section~\ref{sec2} and distributions of target fields derived from observations, we can in principle derive the \gray emissivity profiles. However, the emissivity profiles are derived in 3D space. Meanwhile, the observed profiles are projected on  a 2D surface. The projections are mathematically described by Abel transformation, i.e., the 3D profile $F(R)$ can be transferred to $f(r)$ (hereafter, we use the upper case for 3D values and lower case for 2D values) as 
\begin{equation}
f(r)=2\int_{r}^{R_1}\frac{F(R)R}{\sqrt{R^2-r^2}}dR, 
\label{eq:abel}
\end{equation}
where the upper integration bound $R_1$ should be regarded as the outer radius of the \gray emission region. In what follows, we shall express $R_1$ and $r$ in degrees, which are estimated by dividing the physical distance $R_1$ and $r$ by the distance to the source $d$. 
Such a projection process can be done separately for CRs and target fields.


For CRs, as mentioned above, for continuous injection of CR protons, the 3D profile scale is $F(R) \sim 1/R$. In this case, Eq.\ref{eq:abel} can be integrated analytically, and the result is 

\begin{equation}
f(r)={\rm log}(1+\frac{R_1}{\sqrt{R_1^2-r^2}})-{\rm log}(\frac{R_1}{\sqrt{R_1^2-r^2}}-1).
\label{eq:r1}
\end{equation}
Here, it should be noted that if $R_1$ goes to infinity the integration will diverge; thus, an upper bound of the integration must be set. Physically, such a boundary may come from the limited size of the \gray emission region, caused either by the limited volume of CR protons have occupied or by the distribution of the target fields. 

\begin{figure}[H]
  \centering\includegraphics[width=0.6\linewidth]{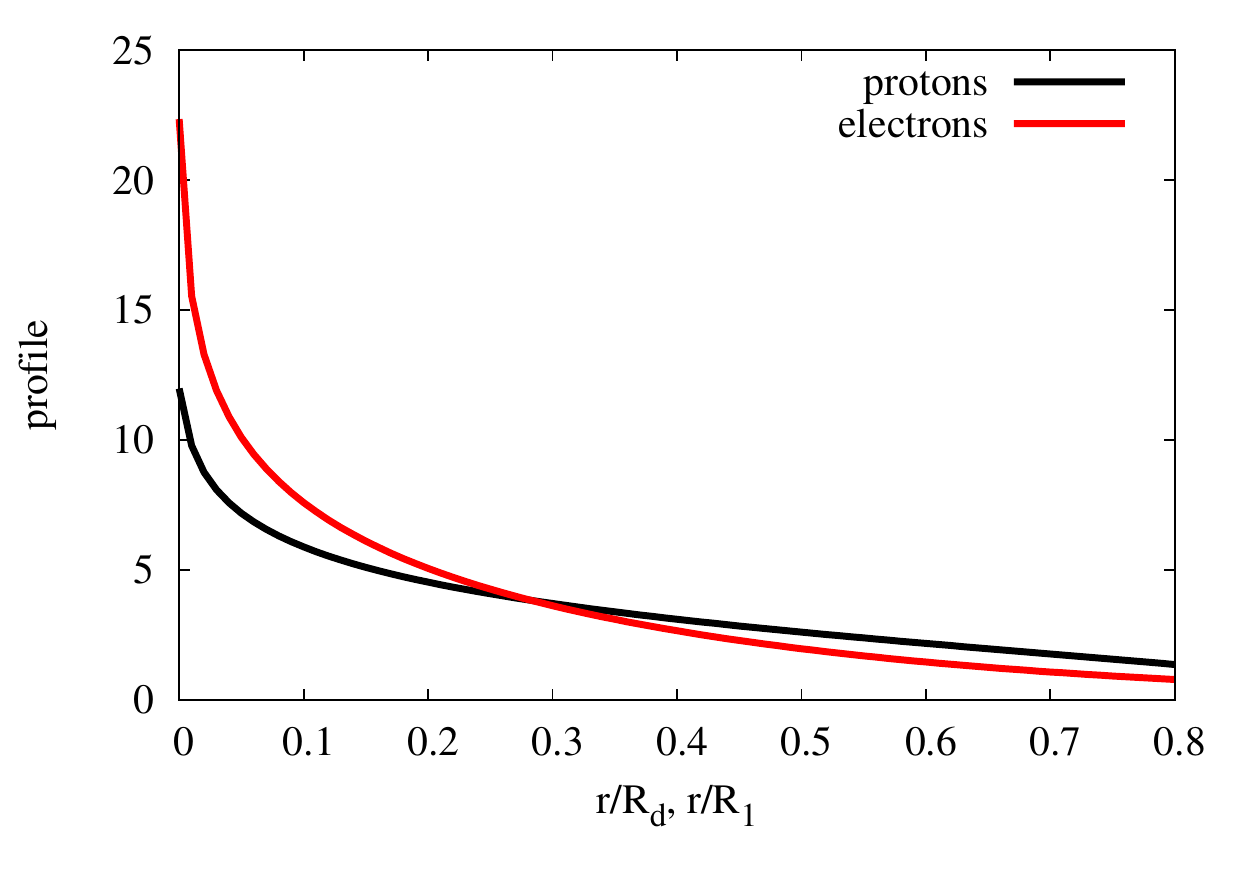}
  \caption{Projected 2D profile (surface brightness) of CR protons and electrons. }
   \label{fig:pro_2d}
\end{figure}

For continuously injected electrons, we have 
\begin{equation}
   F(R) \sim \frac{{\rm erfc}(R/R_{\rm d})}{R}, 
  \label{eq:ele}
\end{equation}
Moreover, this function cannot be integrated analytically; therefore, we choose a numerical approach. The results for protons and electrons are shown in Fig.\ref{fig:pro_2d}. The derived profiles above are for the CRs themselves.
In addition, for the corresponding \gray emissions, the profiles should be convolved with the spatial distribution of the target fields, such as gas and ISRF. 


\section{The influence of the PSF of instruments}
\label{sec4}
\begin{figure*}
  \centering\includegraphics[width=0.45\linewidth]{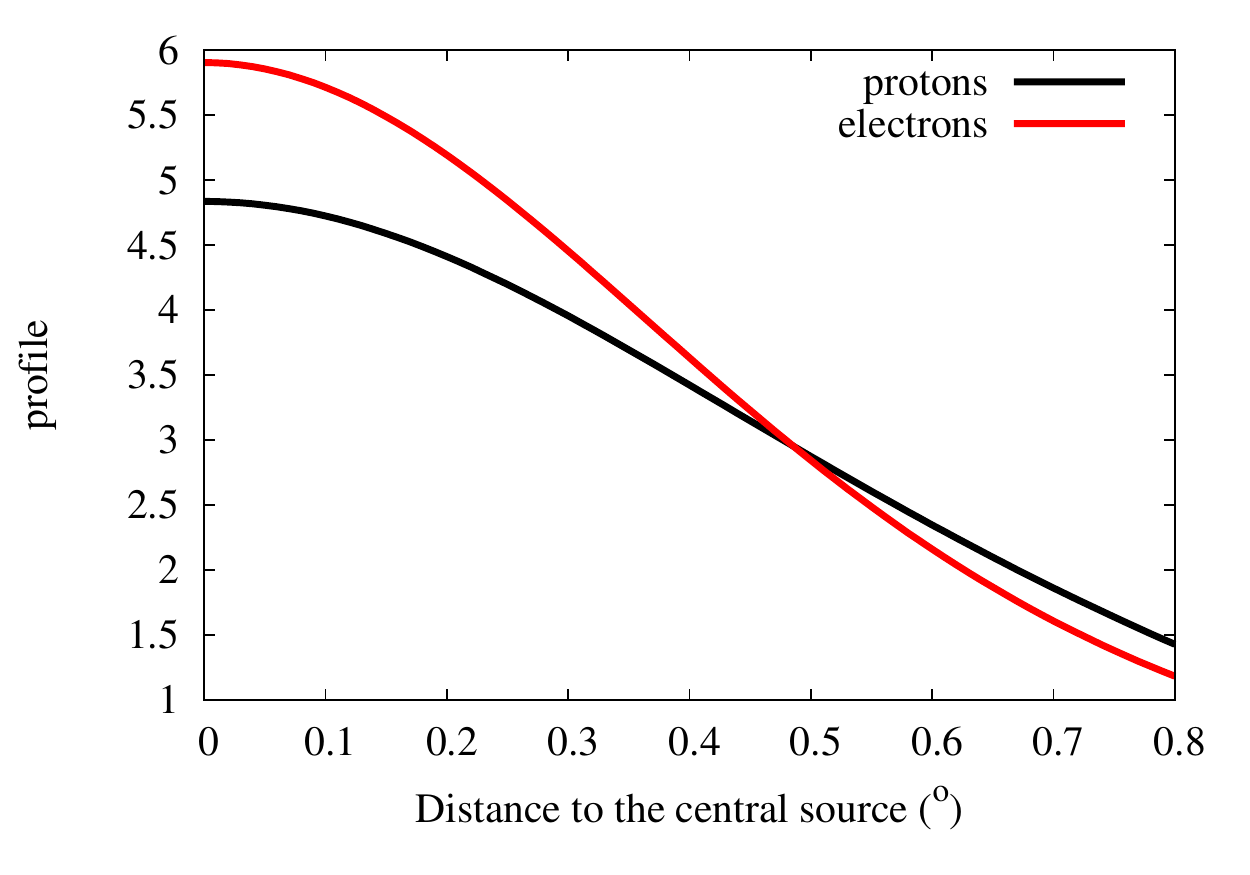}
  \includegraphics[width=0.45\linewidth]{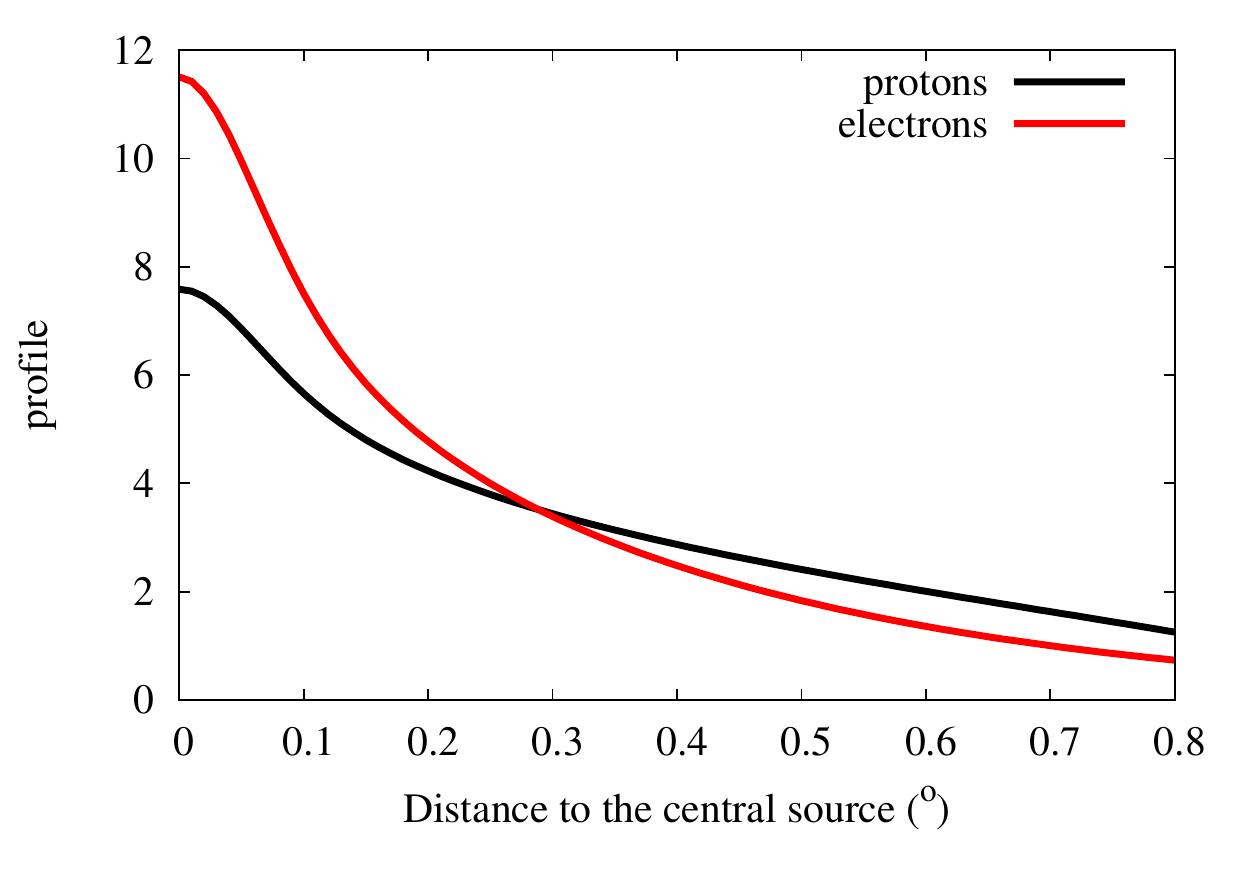}
  \caption{Projected (2D) surface brightness convolved by a Gaussian function with a kernel of $0.3^{\circ}$ (left panel) and  $0.05^{\circ}$ (right panel), respectively. $R_1$ for protons and $R_{\rm d}$ for electrons are both set to be $1^{\circ}$. }
  \label{fig:sm}
\end{figure*}

\begin{figure*}
  \centering\includegraphics[width=0.45\linewidth]{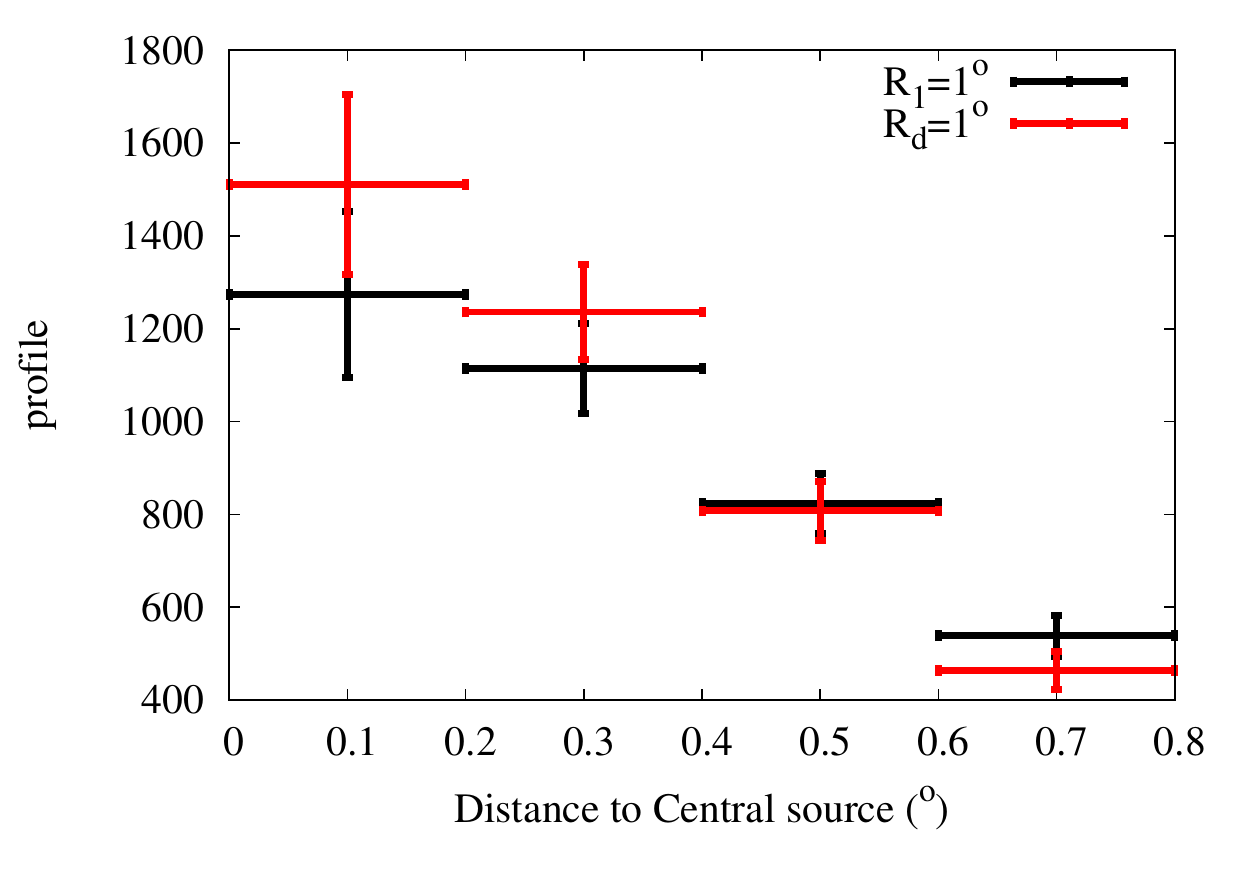}
  \includegraphics[width=0.45\linewidth]{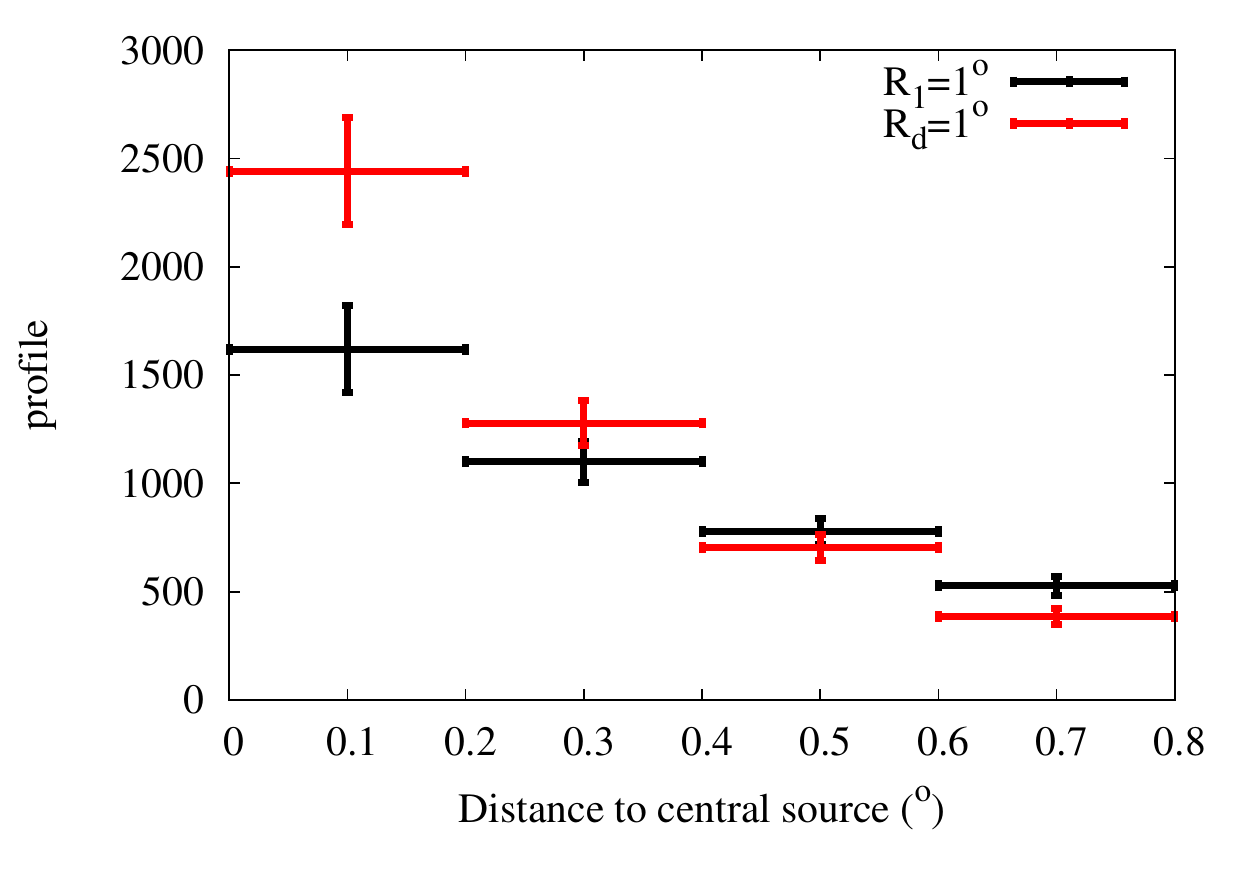}
  \caption{Predicted observed profiles for both electron and proton distributions outside the source, $R_1$ for protons and $R_{\rm d}$ for electrons are both chosen as $1^{\circ}$. The surface brightness is convolved by a Gaussian function with a kernel of $0.3^{\circ}$ (left panel) and  $0.05^{\circ}$ (right panel), respectively. The total photon counts for both cases are assumed to be 500. }
  \label{fig:bin}
\end{figure*}

For realistic observations, the derived projected profile should be convolved with the PSF of the instruments. For \gray astronomy in the GeV band, the PSF of typical instruments such as {\sl Fermi}-LAT is strongly energy-dependent, which is about $1 ^{\circ}$ at 1 GeV and $0.1 ^{\circ}$ above 10 GeV. In the TeV band, there are mainly two types of instruments, the extensive air shower arrays (EASAs) such as ARGO \cite{argo}, HAWC \cite{hawc} and LHAASO \cite{lhaaso2016}, and the imaging air Cherenkov telescope arrays (IACTs) such as H.E.S.S \cite{hess}, Veritas\cite{veritas}, Magic \cite{magic} and CTA \cite{cta}. Generally, the EASAs have an angular resolution of about $0.3 ^{\circ}$ above 10 TeV, whereas for IACTs, the value can be as good as $0.05 ^{\circ}$. To investigate the influence of the PSF, we convolve the \gray profile with the Gaussian function of kernel radii of $0.3 ^{\circ}$ and $0.05 ^{\circ}$, respectively. For simplicity, we assume that  target fields (gas or ISRF) are uniformly distributed; thus, the \gray emission profile should be the same as the projected CR profiles derived in the above section. The results for $R_1=1^{\circ}$ and $R_{\rm d}=1 ^{\circ}$ are shown in Fig.\ref{fig:sm}.
In realistic observations, the statistics are limited. For instance, an exposure of about 200 hours for a source as strong as Crab will collect about 500 photons for a typical IACT with an effective collecting area of $10^5 ~\rm m^2$.  We assume that the total observed photon counts are 500 for the whole region, and the predicted observed profiles are shown in Fig.\ref{fig:bin}. The counts are binned to radial bins with a width of $0.2 ^{\circ}$,  and the statistical error in each bin is estimated as the Poisson noise. Here, we found that it is impossible to distinguish the proton and electron scenarios for the angular resolution of $0.3 ^{\circ}$; however, the difference in the first bin can be as large as $3\sigma$ for the angular resolution of  $0.05 ^{\circ}$. In investigating the profile of \gray surface brightness, the angular resolution of instruments would play a significant role. To  perform such studies, the ideal energy range would be the VHE or even UHE domain (around 100 TeV), where the background is strongly suppressed. In such an energy range, the current IACTs, with a collecting area of $10^5~\rm m^2$ are incapable of collecting enough photons to conduct the morphology studies. LHAASO \cite{he18}, with a collecting area of more than $10^6~\rm m^2$ would be an ideal hunter for UHE sources. However, the limited angular resolution would prevent the solid conclusion from the spatial distribution analyses, as shown in this section. Therefore, the next generation of IACTs such as CTA, with a collecting area of $10^6~\rm m^2$ and angular resolution of $0.05^{\circ}$, would be the ideal instruments to conduct such studies.

\section{Transition from ballistic propagation to diffusion}
\label{sec5}
The propagation regime of CRs depends on the scales. On a scale smaller than the gyro radius of particles, the CRs propagate ballistically.  The transition between the ballistic propagation and diffusion happens approximately at $R_{\rm b} \sim D/c$, where D is the diffusion coefficient, and c is the speed of light. The \gray surface brightness, in this case, is discussed in detail in \cite{prosekin15}. The ballistic propagation has two effects on the resulted \gray surface brightness. First, the static distribution scales as $F(R) \sim 1/R^2$ rather than $1/R$. Second, the CR particles are far from isotropization in this regime, and the observed \grays are only from the particles that propagate toward the observer. As a result, the \gray emission region should be  point-like in the ballistic regime, regardless of the size of target fields and the volume occupied by the CRs. 
Generally, $D$ increases as the energy increases. In other words, $R_{\rm b}$ is small when the energy is low; thus, diffusion dominates in most of the \gray emission region, which resulting to an extended \gray structure. Similarly, $R_{\rm b}$ can be much larger when the energy is high, and in this case, the \gray emission reveals a point-like feature.


Considering a hypothetical extended \gray source with a physical size of 20 parsecs (pc) and a distance of $1~\rm kpc$ (angular size of about $1^{\circ}$), the physical size of the source can be determined by the distribution of gas or the volume occupied by CRs. The $D$ in the Galactic plane is estimated as $4\times 10^{28} (\frac{E}{1\rm GeV})^{\delta}~\rm cm^2/s$ ($0.33<\delta<0.5$) \cite{Strong2007review}. Therefore, for a CR with energy of 1 GeV, $R_{\rm b} \sim 0.4 ~\rm pc$, which is much smaller than the \gray source. In the GeV range, the observed \gray emission should have an extension of about $1^{\circ}$.  For a CR with energy of 1 PeV, $R_{\rm b}$ is already larger than the size of the whole \gray emission regions; thus in this case,  the source should be point-like. It should also be noted that $D$ can be much smaller near the CR sources, so such effects can be minimized. We also note that $R_{\rm b}$ and $R_{\rm d}$ are directly proportional to $D$. Therefore, it is interesting that an exceedingly small or a very large $D$ can result in a "shrink" of the size of the \gray emission.  
The CRs protons and electrons cannot propagate far enough for a very small $D$ since $R_{\rm d}$ scales as $\sqrt{D}$, while for large $D$, the CRs around the source would be in the ballistic regime, and the source will be point-like due to beaming effects.  Recently, such an effect is investigated in detail for Geminga in \cite{recchia21}, in which the \gray profile of Geminga can also be fitted using a Galactic diffusion coefficient by considering the ballistic propagation.

\section{Diffuse emissions}
\label{sec6}
When the surface brightness is very low, the central part of the extended source is invisible. However, if the profile continues as $1/r$ (or more exactly, when $R_{\rm d}$ is large), the diffuse emission may be detected with enough significance when we consider a much larger region around the source. To investigate such a scenario, we assume the distance of the source to be 1 kpc and $R_{\rm d} \sim 200 pc$. The detection significance scales roughly as $N_{\rm src}/\sqrt{N_{\rm bg}}$, can be calculated by integrating the projected profile (combining Eq.\ref{eq:abel} and Eq.\ref{eq:ele}), where $N_{\rm src}$ is the total counts from the extended structure. For simplicity, we assume that the background is uniform, thus $N_{\rm bg} \propto A$,
where $A$ is the total area of the integrated area. Thus, the significance $S$ within the integration radius $R_{\rm int}$ can be estimated as 
\begin{eqnarray}
S(R_{\rm int}) &\sim& \frac{1}{\sqrt{A}}\int_{0}^{R_{\rm int}}f(r)rdr \nonumber \\  &\sim& \frac{1}{R_{\rm int}}\int_{0}^{R_{\rm int}}\int_{r}\frac{erfc(R/R_{\rm d})}{\sqrt{R^2-r^2}}dRrdr 
\label{eq:sig_r}    
\end{eqnarray}

\begin{figure*}
  \centering\includegraphics[width=0.45\linewidth]{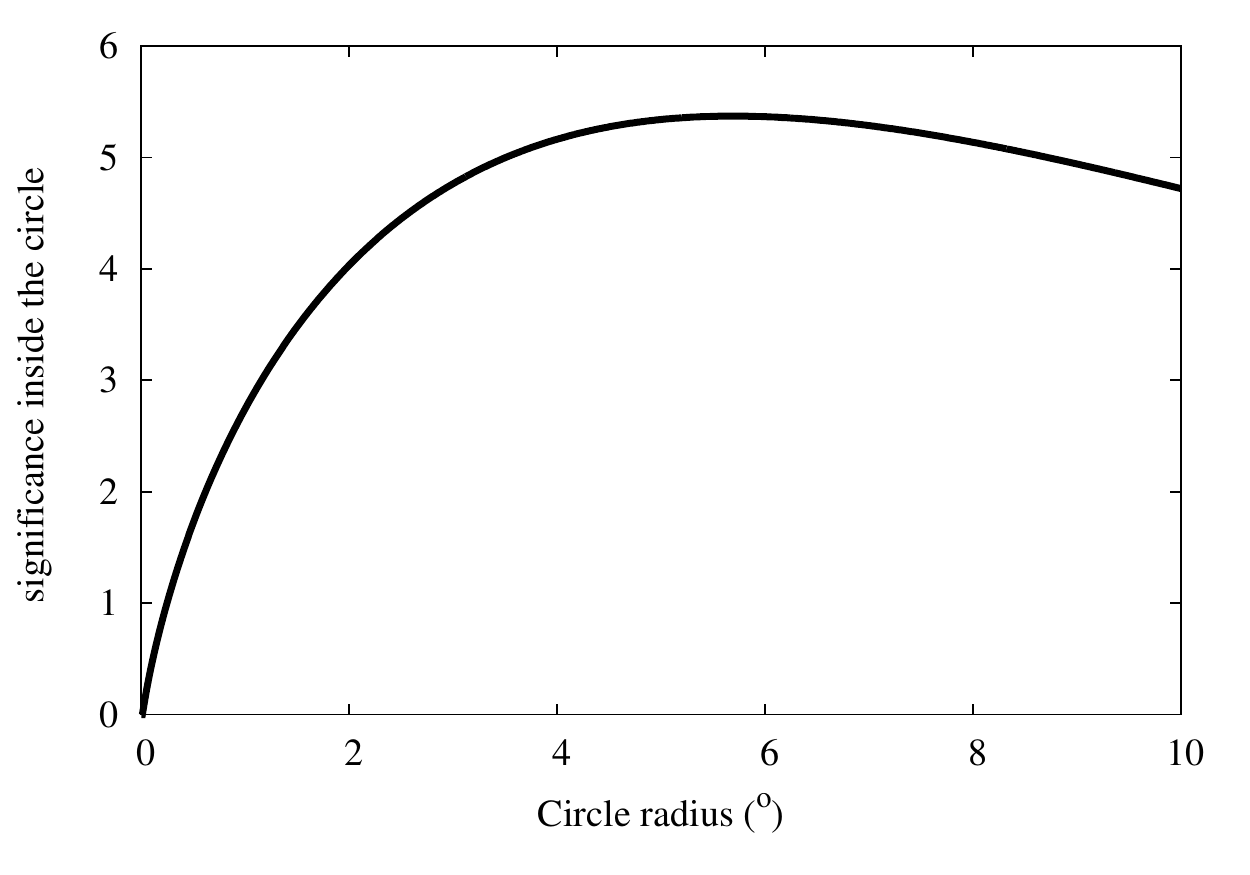}
  \includegraphics[width=0.45\linewidth]{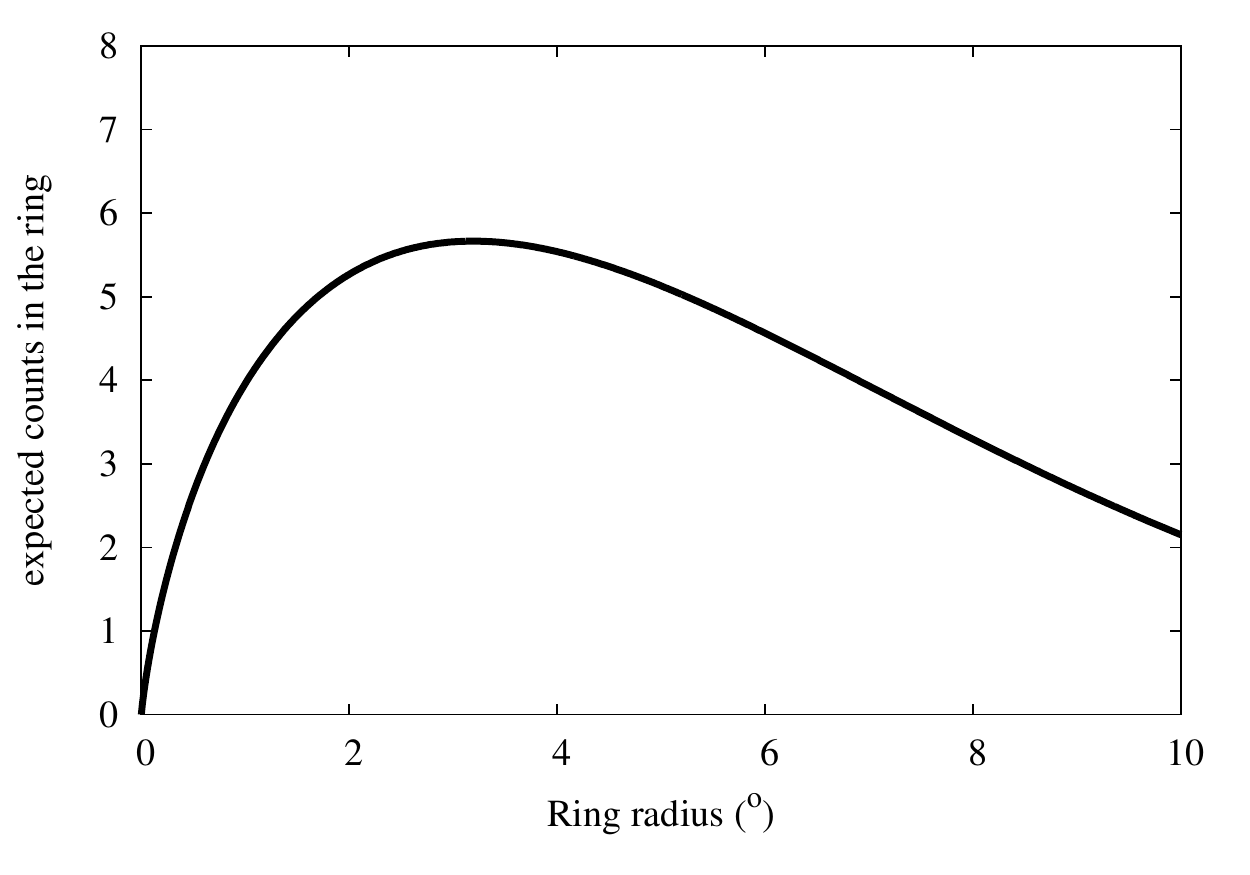}
  \caption{{\it Left panel}: Detection significance for different integration radius around the source center; the normalization is arbitrary. {\it Right panel}: Expected photon counts in the rings with the same thickness but  different distances to the source center; the normalization is arbitrary. }
  \label{fig:R_dif}
\end{figure*}

The results (with arbitrary normalizations) are shown in the left panel of Fig.\ref{fig:R_dif}. We found that the significance increases quickly when we increase the integrated radius and reaches the maximum at about $5^{\circ}$. We also estimated the probability of detecting a photon at each distance from the source center (where CRs are injected), which is done by calculating the expected photon counts in each ring ($r'$, $r'+dr'$) with different distances to the source center, as expressed below.
\begin{equation}
N(r') \sim 2\pi \int_{r'}^{r'+dr'}f(r)rdr=4\pi\int_{r'}^{r'+dr'}\int_{r}\frac{erfc(R/R_{\rm d})}{\sqrt{R^2-r^2}}dR r  dr
\label{eq:dis_r}    
\end{equation}

The results (with arbitrary normalizations)  are shown in the right panel of Fig.\ref{fig:R_dif}. For the parameters used here the  photon counts peaks at  $r' \sim 3 ^{\circ}$, so in case of a very low total flux, the first several photons may be detected at the regions quite far from the center, and such an extended source may be identified as diffuse emission. Recently, the Tibet $\rm AS\gamma$ collaboration claimed the detection of diffuse \gray emission up to more than 900 TeV \cite{asgamma21}. In their studies, the inner $0.5^{\circ}$ are masked out near the known source. As we shown here, for such halo or cocoon like sources, the photons may be first detected much further away from the source. Therefore, a $0.5^{\circ}$ mask may be insufficient to rule out all the contributions from the sources. 
In the energy range above 100 TeV, the \gray observations may be operated in a background free regime \citep{lhaaso_nature}, in this ideal case without background, the Poisson statistics would not apply, and the detection significance of the source is labeled as the detected photon counts, which is simply the integration of Eq.\ref{eq:dis_r}. The derived significance should increase monotonically,  and when the photon counts are small the source will be regarded as the true diffuse emission rather than an extended source.

\section{Discussions} 
\label{sec7}
HAWC collaboration has discovered  an extended halo structure near the Geminga pulsar \cite{hawc_geminga}. In the energy range of HAWC, only cosmic microwave background  is responsible for the inverse Compton (IC) emissions; thus, the \gray distribution should be the same as the electron distribution. The angular size of the halo is more than $5^{\circ}$; the profile was fitted with a projected IC surface brightness profile considering the diffusion and cooling of electrons. The best-fit $R_{\rm d}$ reveals a diffusion coefficient of two orders of magnitude less than the average value in the Galactic plane. H.E.S.S collaboration measured the TeV \gray profile near the GC and derived a $1/r$ profile of CR protons by considering the gas distribution. This result implies a continuous injection of CRs from the compact region of GC. Such $1/r$ CR proton profiles are further found in the surroundings of several YMCs, such as Cygnus OB2 and Westerlund~1, which indicates that the YMCs can provide significant contributions to the Galactic CRs. These studies show the power of the accurate measurement of \gray radial profiles. 

\begin{figure}
  \centering\includegraphics[width=0.7\linewidth]{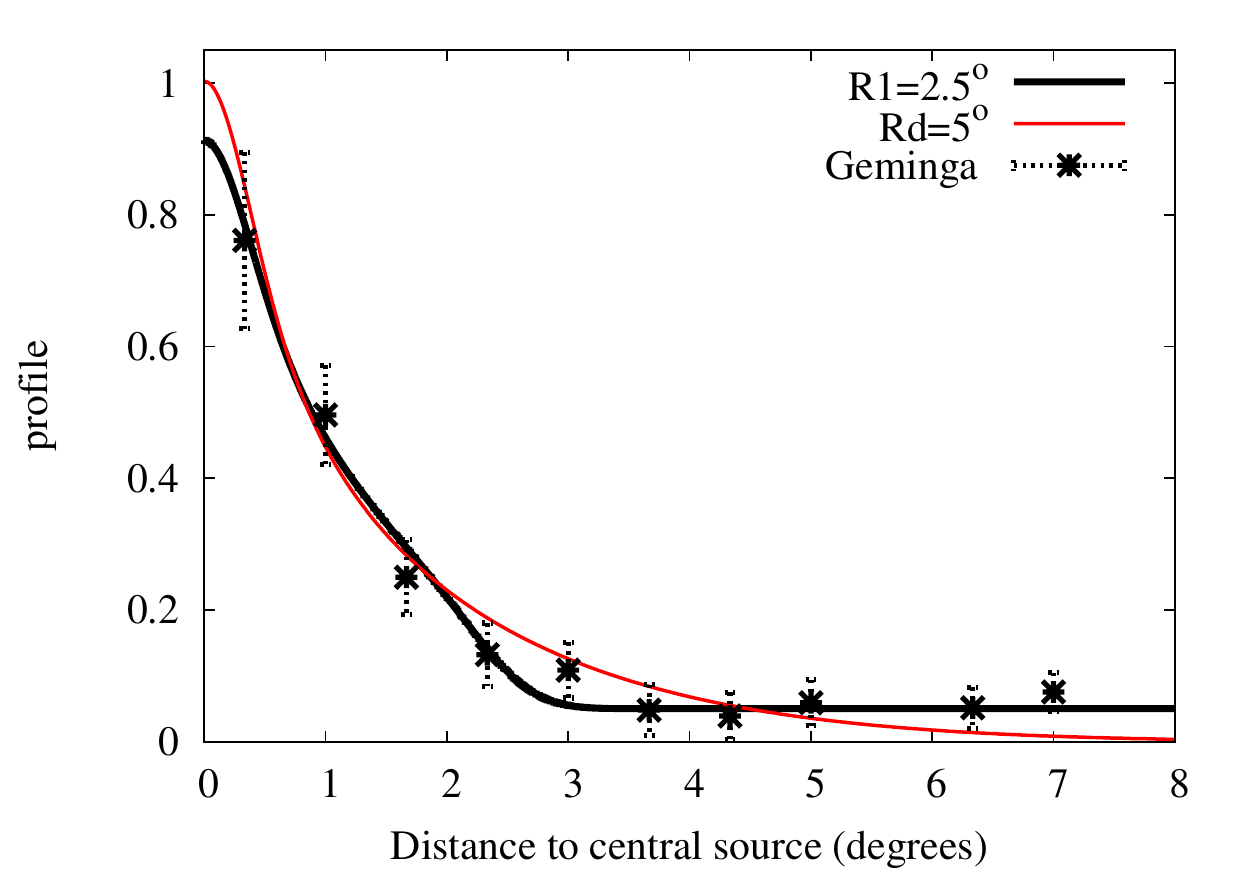}
  \caption{\gray surface brightness profile of Geminga observed by HAWC \cite{hawc_geminga}. The red curve reveals the projected electron distribution with $R_{\rm d} = 5.0^{\circ}$ and the black curve is the projected proton distribution with $R_1 = 2.5^{\circ}$. Both curves are smoothed by a Gaussian function with a kernel of $0.3^{\circ}$.  }
  \label{fig:geminga}
\end{figure}

As an example of our former calculations, we revisited the fitting of the Geminga profile  using the analytical 3D electron distributions and numerical projection to the 2D surface and convolved with a Gaussian function with a kernel of $0.3^{\circ}$. The results are shown with a  red curve in Fig.\ref{fig:geminga}. The derived best-fit $R_{\rm d}$ is $5.0^{\circ}$, which is consistent with the results in \cite{hawc_geminga}. However, we found the same profile can also  be fitted with the projected $1/r$ profile with a smaller $R_1$ of about $2.5^{\circ}$. We note that in the fitting of $1/r$ profile, an additional uniform background is added to fit the flat distribution beyond $3.0^{\circ}$ from Geminga. But such a uniform background is quite reasonable for TeV \grays due to the CR contamination. Thus, the profile itself cannot rule out the hadronic origin of the \gray emission. Of course, for this case, the leptonic origin is much more natural considering the presence of the pulsar and the lack of CR proton accelerators in the vicinity. A hadronic origin requires an unrealistic energy budget in parent particles \citep{liu19,bao2021arXiv}. 

\begin{figure}
  \centering\includegraphics[width=0.7\linewidth]{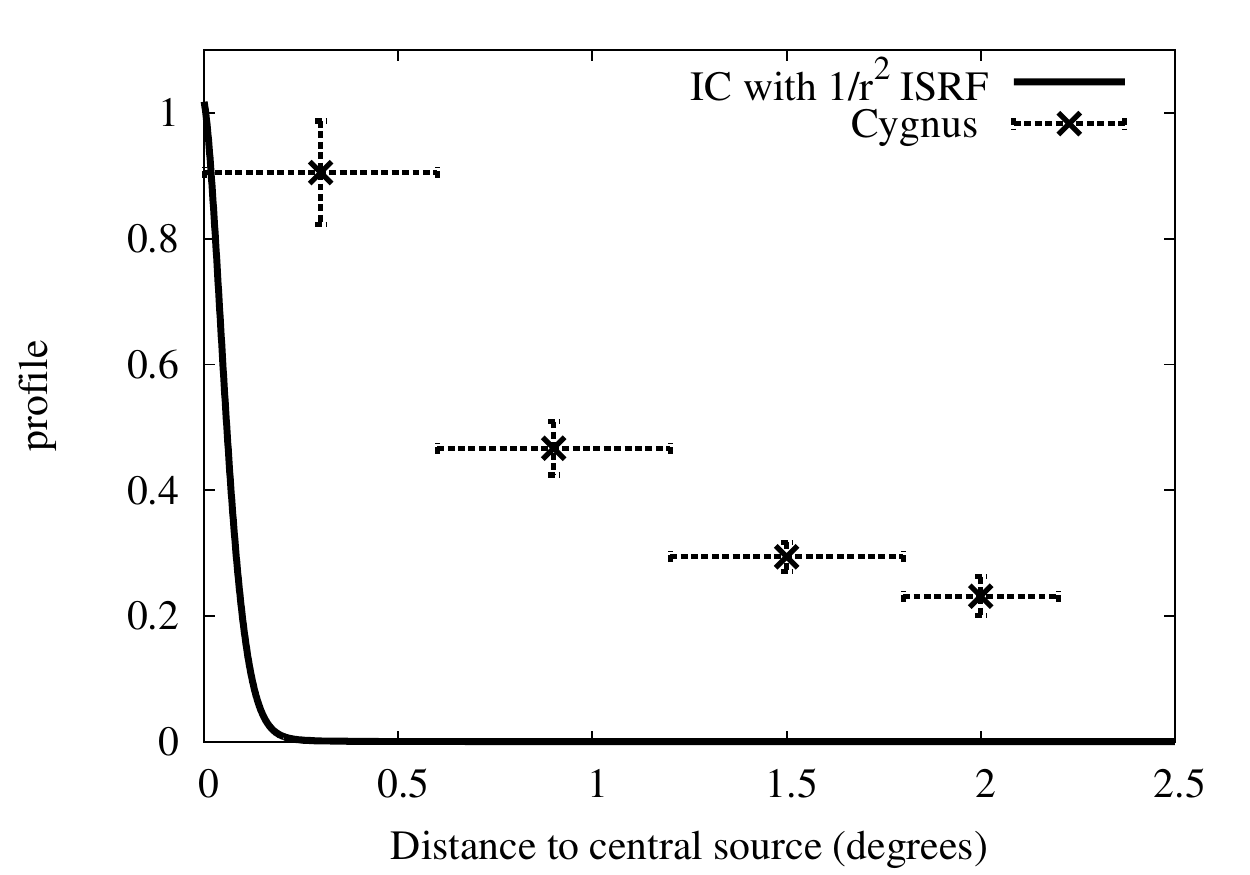}
  \caption{\gray  profile of Cygnus cocoon derived in  \cite{aharonian19}.  The black curve is projected IC profile assuming a $1/r^2$ ISRF distributions, which is smoothed by a Gaussian function with a kernel of $0.05^{\circ}$.  }
  \label{fig:cygnus}
\end{figure}

Furthermore, the same test can also apply to the CR cocoons near GC and YMCs, which are identified as CR proton accelerators due to the $1/r$ structure. Indeed, the projected $1/r$ profile should be well fitted with the projected electron diffusion profile with a larger $R_{\rm d}$ than $R_1$ as shown in Fig.\ref{fig:geminga}. However, for the GC and YMCs, the projected $1/r$ profiles are derived by dividing the gas distributions. Unlike the case for Geminga, where the ISRF is uniform, the ISRF near GC and YMCs are significantly enhanced near the central source due to the starlight from the massive stars. The  starlight component  of  ISRF can be many orders of magnitude higher within several pc of the central sources and scale as $1/r^2$. Such enhancement should have two effects on the \gray profile. First, the distribution of electrons should be modified since the cooling is also spatial dependent. Second, the electron distributions should be convolved with the $1/r^2$ profile of the ISRF to get the \gray distributions. For the reasonable sizes and luminosities of YMCs, the first effect is negligible since the ISRF drops very fast (as $1/r^2$) and after several pc the average ISRF in the Galactic plane should dominate, and the cooling effect is again spatially independent. However, the second effect has a significant impact on the \gray profile. As an example, we plot the predicted IC \gray profile of Cygnus cocoon assuming a leptonic origin in Fig.\ref{fig:cygnus} and a $1/r^2$ distribution of the ISRF. In this case, we found that the derived profile is much sharper than the observed profile, and it cannot be distinguished from a point source, even for a PSF of $0.05^{\circ}$. Indeed,  a bright central source is found for both GC and Cygnus cocoon. In the previous study, it is explained as an additional and irrelevant source, such as pulsar wind nebula. This is a rational explanation of such kind of systems, considering their crowded nature. However, it is also possible that the central \gray point source is caused by the CR electrons injected together with CR protons that fill the $1/r$  CR proton "cocoon" as we have demonstrated.  It should also be noted that due to the Klein-Nishina effect, the starlight component of ISRF would be negligible for IC emissions above 1 TeV; thus, the corresponding corresponding central \gray point source should have a spectral cutoff at about TeV energy, which will be tested in future observations.

Recently, the LHAASO-KM2A array has revealed a population of UHE \gray sources, most of which are extended \cite{lhaaso_nature}. Several of these sources are located in the vicinity of pulsars or YMCs and would be promising candidates for CR cocoons and TeV halos. As mentioned in Section~\ref{sec4}, the limited angular resolution of EASAs prevents the separation of different profiles. Moreover, the current IACTs can hardly collect enough photon counts to study the radial profile due to the limited collecting area. Therefore, forthcoming  IACTs, such as CTA, with an effective collecting area of  $10^6~\rm m^2$ and a better angular resolution $<0.05^{\circ}$, would be the ideal instruments to  perform such kind of investigation and provide us with unique information on the distribution and propagation of relativistic particles and shed light on the origin of CRs in our Galaxy.

\vspace*{2mm} \Acknowledgements{\bahao Ruizhi Yang is supported by  the NSFC under grants 11421303, 12041305 and  the national youth thousand talents program in China. Bing Liu is supported by the Fundamental Research Funds for the Central Universities. }






\end{document}